# A Novel Provably Secure Key-Agreement Using Secret Subgroup Generator

Abdelhaliem Babiker

College of Engineering, Imam Abdurrahman Bin Faisal University

Email: aababiker@iau.edu.sa; haliem.abbas@gmail.com

**Abstract-** In this paper, a new key-agreement scheme is proposed and analyzed. In addition to being provably secure in shared secret key indistinguishability model, the scheme has an interesting feature: while using exponentiation over a cyclic subgroup to establish the key-agreement, the generator of that subgroup is hidden to secure the scheme against adversaries that are capable of solving the Discrete Logarithm Problem, which means that the scheme might be candidate as a post-quantum key exchange scheme.

## 1.1 Introduction

Theoretic and technical developments the field of the Quantum Computers− a computers based on principles of quantum physics and utilizes these principles to solve problems that would require infinitely huge amount of time and computational capacity using classical computers− are imposing real challenge on the field of public-key cryptography; most of the currently hard or intractable problems which forms the basis for security of the widely used cryptographic algorithms such as Discrete Logarithm Problem and Integer Factorization Problem will be solved using large-scale quantum computers. Consequently, many of the currently used public-key cryptosystems will be compromised [1]. Furthermore, the current cryptosystems which are based number theory are becoming less secure in light of the developments in mathematical and computational cryptanalysis [2][3]. Therefore, there is great interest in developing new cryptographic algorithms that would be secure against both quantum and classical computers.

In this paper we propose new key-agreement scheme that uses exponentiations over subgroup of square matrices over $\mathbb{F}_2$, and yet− unlike many other exponentiation-based key agreements and cryptosystems such as Diffie-Hellman key exchange and ElGamal cryptosystem− does not assume intractability of the Discrete Logarithm Problem.

The Discrete Logarithm Problem (DLP) can generally be stated as follows. Given $g$ an element in some cyclic group $G$, and $g^\alpha$ for an integer $\alpha$; find $\alpha$. When the generator $g$ is an unknown, we would have an Unknown-Base DLP which is one equation with two unknowns, and it has at least, $p$ possible solutions in any cyclic group of order $p$.

The basic idea in the proposed scheme is a simple one, we hide the actual group generator being used as base for the exponentiations in the key agreement to deprive the adversary from any advantage of solving the DLP. By hiding the actual subgroup generator, we get rid of the reliance on intractability of the DLP such that capability of solving the DLP does not mean breaking the scheme.

However, for any key agreement protocol to surpass the level of security provided by intractability of the DLP, it has to reach it first. That is, the protocol must first be provably secure under standard intractability assumptions relevant to the DLP. Therefore, at this level, security of scheme is proved using key indistinguishability model, by showing that the shared secret key is indistinguishable from the random under Decisional Diffie-Hellman (DDH) assumption for subgroup of matrices over $\mathbb{F}_2$ with prime multiplicative order.

One can easily note that the DDH assumption used in proof of indistinguishability of the shared is reducible to the DLP. This implies that indistinguishability proof would not be valid when the DLP is broken. However, the DDH assumption is used proof of the security against classical (non-quantum) adversary. And it is presented to show that the proposed scheme satisfies basic security standards.

Regarding an adversaries that are capable of breaking the DLP, such as quantum adversaries, indistinguishability of the shared secret key not guaranteed. But the paper shows that solving DLPs derived from the scheme's security equations would not imply computing the shared secret key.

*Contribution of this paper*

This key exchange scheme gets rid of the reliance on the DLP while retaining simple use of exponentiations to establish key agreement, which means it might be a candidate as a post-quantum key exchange scheme. Furthermore, the scheme might also be applicable in different non-commutative platforms that have the appropriate structural properties.

The paper is organized as follows. In section 2 we introduce the key agreement scheme. In section 3 we give a basic proof for security of scheme against non-quantum adversary using key indistinguishability model. Section 4 discuss security of the scheme against a quantum adversary (or any hypothetical adversary) who is able to solve any kind of DLP, showing that the shared secret key will remain secure and hidden from such adversary. Then conclusion is given in section 5.

## 2 Preliminary

A *Binary matrix B* is a matrix over the binary field $\mathbb{F}_2$ ( i.e. $B_{n \times n}$: $B_{ij} \in \{0,1\}$). In the context of this paper, matrix always refers to nonsingular binary matrix, also multiplication and exponentiation, whenever appears, refers to binary matrix multiplication and binary matrix exponentiation where arithmetic operations are performed modulo 2.

## 3 The Key Agreement

In this protocol Alice and Bob first agree on prime number $p$ (generated by algorithm 2.3) then the key agreement goes as follows.

(1) Bob selects there *t-tuples* of positive integers $\boldsymbol{\mu}, \boldsymbol{\sigma}$ and $\boldsymbol{\theta} = (\theta_1, \ldots, \theta_t)$ (using algorithm 2.1) then sends $(\boldsymbol{\mu}, \boldsymbol{\sigma})$ to Alice, and keeps $\boldsymbol{\theta}$ privately.
(2) Alice generates secret random matrices $R$ and $S$ such that $RS \neq SR$, selects secret numbers $\alpha, \gamma$, generates two *t-tuples* of matrices $\boldsymbol{A} = (A_1, \ldots A_t)$, $\boldsymbol{B} = (B_1, \ldots B_t)$ (using algorithm 2.2), then sends $(\boldsymbol{A}, \boldsymbol{B})$ to Bob.
(3) Bob computes his secret key $K = B_1^{\theta_1} \ldots B_t^{\theta_t}$, and sends $Y = A_1^{\theta_1} \ldots A_t^{\theta_t}$ to Alice.
(4) Alice computes her secret key $K = SR^{-1}Y^{\gamma\alpha^{-1}}RS^{-1}$.

The shared secret key is $K$.

The key exchange protocol is fully described by the distribution

$$\mathcal{D} = (\boldsymbol{\mu}, \boldsymbol{\sigma}, \boldsymbol{A}, \boldsymbol{B}, Y, K)$$

**Algorithm 2.1**
Input: $\phi = 6p$ for some prime $p$, positive integer $t$.
Output: *t-tuples* of positive integers $\boldsymbol{\sigma}$, $\boldsymbol{\mu}$, $\boldsymbol{\theta}$.

(1) Select positive integers from $[1, p]$, $\sigma_1, \ldots, \sigma_{t-1}, \mu_1, \ldots, \mu_{t-1}$, and $\theta_1, \ldots, \theta_{t-1}$ such that $\sigma_i \bmod 3 \neq 0$, $\mu_i \bmod 2 \neq 0$, $\gcd(\phi, 3p - \sum_{i=1}^{t-1} \sigma_i \theta_i) = 1$ and $\gcd(\phi, 2p - \sum_{i=1}^{t-1} \mu_i \theta_i) = 1$.
(2) Select positive integer $\theta_t$ such that $\gcd(\phi, \theta_t) = 1$, then compute
$\sigma_t = \theta_t^{-1}(3p - \sum_{i=1}^{t-1} \sigma_i \theta_i) \bmod 3p$, and $\mu_t = \theta_t^{-1}(2p - \sum_{i=1}^{t-1} \sigma_i \theta_i) \bmod 2p$.
(3) Return $\boldsymbol{\sigma} = (\sigma_1, \ldots, \sigma_t)$, $\boldsymbol{\mu} = (\mu_1, \ldots, \mu_t)$, $\boldsymbol{\theta} = (\theta_1, \ldots, \theta_t)$.

Note that $(\mu_1 \theta_1 + \cdots + \mu_t \theta_t) \equiv 0 \bmod 2p$, and $(\sigma_1 \theta_1 + \cdots + \sigma_t \theta_t) \equiv 0 \bmod 3p$

**Algorithm 2.2**
Input : *t-tuples* of integers $\boldsymbol{\mu}, \boldsymbol{\sigma}$, secret numbers $\alpha, \gamma$ and secret matrices $R, S$.
Output: two *t-tuples* of matrices $\boldsymbol{A} = (A_1, \ldots A_t)$, $\boldsymbol{B} = (B_1, \ldots B_t)$.

(1) Generate secret $M$ matrix with multiplicative order $\phi = 6p$, for some prime $p$ (using algorithm 2.3). Compute $T = M^6$, $U = M^3$, $V = M^2$.
(2) Select $\zeta_1, \ldots, \zeta_t \in [1, p]$ such that $(2\alpha\zeta_i + \mu_i) \bmod p \neq 0$ and $(3\gamma\zeta_i + \sigma_i) \bmod p \neq 0$.
(3) Compute $A'_1 = T^{\alpha\zeta_1} U^{\mu_1}$, $A'_2 = T^{\alpha\zeta_2} U^{\mu_2}$, ..., $A'_t = T^{\alpha\zeta_t} U^{\mu_t}$.
(4) Compute $B'_1 = T^{\gamma\zeta_1} V^{\sigma_1}$, $B'_2 = T^{\gamma\zeta_2} V^{\sigma_2}$, .., $B'_t = T^{\gamma\zeta_t} V^{\sigma_t}$.
(5) Select two random secret matrices $R$ and $S$ such that $RS \neq SR$.
(6) Set $A_i = R A'_i R^{-1}$, and $B_i = S B'_i S^{-1}$ ; $i = 1, \ldots, t$.
(7) Return $\boldsymbol{A} = (A_1, \ldots A_t)$ and $\boldsymbol{B} = (B_1, \ldots B_t)$.

**Algorithm 2.3**
(1) Obtain a primitive polynomial $P(x)$ of degree $m$, such that $2^m - 1$ is prime $p$, or has large prime factor $p$.
(2) Construct the companion matrix $C$ of $P(x)$.
(3) Generate random $n \times n$ matrix $B$ ($n = m + 3$).
(4) Construct $M = BNB^{-1}$; where $N = \begin{bmatrix} D & \\ & C^{(2^m-1)/p} \end{bmatrix}$; $D$ is $3 \times 3$ such that $D^6 = I_3$.
(5) Return $M$.

Note that if $2^m - 1$ is prime $p$ then $C^{(2^m-1)/p} = C$.

Now, the primitive polynomial $P(x)$ is minimal polynomial of its companion matrix $C$, therefore the multiplicative order of $C$ is $2^m - 1$.
Since $D^6 = I_6$, then $M^{6p} = B \begin{bmatrix} D^{6p} & \\ & C^{6(2^m-1)} \end{bmatrix} B^{-1} = I_n$.
Thus, the multiplicative order of $M$ is $\phi = 6p$.

**Validity of The key Agreement.**

*Alice's Key*
$Y = (A_1^{\theta_1} \ldots A_t^{\theta_t}) = R(A'_1{}^{\theta_1} \ldots A'_t{}^{\theta_t}) R^{-1}$.
$= R(T^{\alpha\zeta_1} U^{\mu_1})^{\theta_1} (T^{\alpha\zeta_2} U^{\mu_2})^{\theta_2} \ldots (T^{\alpha\zeta_t} U^{\mu_t})^{\theta_t} R^{-1}$.
$= R(T^{\alpha\zeta_1\theta_1} T^{\alpha\zeta_2\theta_2} \ldots T^{\alpha\zeta_t\theta_t}) (U^{\mu_1\theta_1} U^{\mu_2\theta_2} \ldots U^{\mu_t\theta_t}) R^{-1}$.
$= R T^{\alpha \Sigma \zeta_i \theta_i} U^{\Sigma \mu_i \theta_i} R^{-1}$.
$= R T^{\alpha \Sigma \zeta_i \theta_i} R^{-1}$.     (Since $\Sigma \mu_i \theta_i \equiv 0 \bmod 2p$; $2p$ is multiplicative order of $U$)
$Y^{\gamma\alpha^{-1}} = R(T^{\alpha \Sigma \zeta_i \theta_i})^{\gamma\alpha^{-1}} R^{-1} = R T^{\gamma \Sigma \zeta_i \theta_i} R^{-1}$.
$K = S R^{-1} Y^{\gamma\alpha^{-1}} R S^{-1} = S T^{\gamma \Sigma \zeta_i \theta_i} S^{-1}$.

*Bob's Key*

$$K = (B_1^{\theta_1} \ldots B_t^{\theta_t}) = S(B'^{\theta_1}_1 \ldots B'^{\theta_t}_t)S^{-1}.$$
$$= S(T^{\gamma\zeta_1}V^{\sigma_1})^{\theta_1}(T^{\gamma\zeta_2}V^{\sigma_2})^{\theta_2}\ldots(T^{\gamma\zeta_t}V^{\sigma_t})^{\theta_t}S^{-1}.$$
$$= S(T^{\gamma\zeta_1\theta_1}T^{\gamma\zeta_2\theta_2}\ldots T^{\gamma\zeta_t\theta_t})(V^{\sigma_1\theta_1}V^{\sigma_2\theta_2}\ldots V^{\sigma_t\theta_t})S^{-1}.$$
$$= ST^{\gamma\Sigma\zeta_i\theta_i}V^{\Sigma\sigma_i\theta_i}S^{-1}.$$
$$= ST^{\gamma\Sigma\zeta_i\theta_i}S^{-1}. \text{ (Since } \Sigma\sigma_i\theta_i \equiv 0 \bmod 3p; 3p \text{ is multiplicative order of } V)$$

## 4 Security Against Non-Quantum Adversary

Every instance of the key agreement is fully identified by $\mathcal{D} = (\boldsymbol{\mu}, \boldsymbol{\sigma}, \boldsymbol{A}, \boldsymbol{B}, Y, K)$. $K$ is the shared secret key.

The proof presented in this section is a basic proof, against non-quantum adversary, in order to show that the protocol is secure under a standard intractability assumption, namely Decisional Diffie-Hellman assumption. In the next section, we will give security analysis against quantum adversary who can break the DLP, showing that the shared secret key will remain secure even though. In case of the quantum adversary, the proof based on intractability of the Decisional Diffie-Hellman or the DLP will, of course, be invalid. However, the analysis in section 4 shows that the best of what quantum adversary could have from the solving all possible DLPs derived from security equations of the protocol, is a system of linear equations with number of unknowns greater than number of the equations. Therefore, the adversary will not be able to compute the shared secret key from these equations.

In this section we are going to prove security of the scheme assuming that the matrices $M$ (generated by algorithm 2.3), and $R, S$ (generated by algorithm 2.2) are public. In the actual protocol in the reality, the matrices $M, R,$ and $S$ are private. Thus, if the protocol is secure when these matrices are public, it must be secure when these matrices are private.

We use key indistinguishability model of security proof showing that there is no probabilistic polynomial time algorithm $\mathcal{A}$ to distinguish between the actual secret key $K = ST^{\gamma\Sigma\zeta_i\theta_i}S^{-1}$ and random matrix $K^* = ST^cS^{-1}$ for some integer $c$, under Decisional Diffie-Hellman assumption for subgroup of matrices over $\mathbb{F}_2$ with prime order.

### 4.1 Proof of Indistinguishability of The Shared Secret Key

In what follows we prove indistinguishability of the secret key $K$ from the random matrix $K^*$ using DDH assumption for the subgroup $\mathbb{G}$ of matrices over $\mathbb{F}_2$ generated by matrix $B$. This assumption can be stated as follows.

Let $\mathcal{G}_p$ be family of subgroups of $n \times n$ matrices over $\mathbb{F}_2$ of prime order $p$. For any $\mathbb{G} =<B> \in \mathcal{G}_p$, the proper Diffie-Hellman quadruple $(B, B^a, B^b, B^{ab})$ is indistinguishable from the quadruple $(B, B^a, B^b, B^c)$ where $a, b$ and $c$ ($\neq ab$) are random integers in $(1, p)$.

The assumption can be stated formally as follows [4]. There is no probabilistic polynomial time algorithm $\mathcal{A}$ such that

$$|Pr[\mathcal{A}(B, B^a, B^b, B^{ab}) = 1] - Pr[\mathcal{A}(B, B^a, B^b, B^c) = 1]| > \frac{1}{n^\epsilon}$$

for some $\epsilon > 0$ and sufficiently large $n$. Where $\mathcal{A}(...) = 1$ means the algorithm returns 1 when the quadruple input is proper Diffie-Hellman quadruple.

**Notation**

Let $Y' = R^{-1}YR$, $K' = S^{-1}KS$. Thus, $Y' = T^{\alpha \sum \zeta_i \theta_i}$ and $K' = T^{\gamma \sum \zeta_i \theta_i}$.

Let $\boldsymbol{Q} = (T^{\alpha \zeta_1}, Y', T^{\gamma \zeta_1}, K') = (T^{\alpha \zeta_1}, (T^{\alpha \zeta_1})^a, (T^{\alpha \zeta_1})^b, (T^{\alpha \zeta_1})^{ab})$. $\boldsymbol{Q}$ is Diffie-Hellman quadruple, since $Y' = (T^{\alpha \zeta_1})^a$, $T^{\gamma \zeta_1} = (T^{\alpha \zeta_1})^b$, $K' = Y'^b$; where $a = \zeta_1^{-1} \Sigma \zeta_i \theta_i$, $b = \gamma \alpha^{-1}$. $i \in [1, t]$.

Let $\boldsymbol{Q}^* = (T^{\alpha \zeta_1}, Y', T^{\gamma \zeta_1}, T^c)$, and $\boldsymbol{\mathcal{D}}^* = (\boldsymbol{\mu}, \boldsymbol{\sigma}, \boldsymbol{A}, \boldsymbol{B}, Y, K^*)$.

Thus, $\boldsymbol{Q}$ is indistinguishable $\boldsymbol{Q}^*$ under DDH assumption for the subgroup $\mathbb{G}$ of matrices over $\mathbb{F}_2$.

**Theorem 1**

The distributions $\boldsymbol{\mathcal{D}} = (\boldsymbol{\mu}, \boldsymbol{\sigma}, \boldsymbol{A}, \boldsymbol{B}, Y, K)$ that describes a typical instance of the key agreement protocol and $\boldsymbol{\mathcal{D}}^* = (\boldsymbol{\mu}, \boldsymbol{\sigma}, \boldsymbol{A}, \boldsymbol{B}, Y, K^*)$ are indistinguishable under DDH assumption for the subgroup $\mathbb{G}$ of matrices over $\mathbb{F}_2$. Therefore, the secret key $K$ is indistinguishable from the random $K^*$ under this assumption.

**Proof**

We are going to prove that if there is a polynomial time algorithm $\mathcal{B}$ that distinguishes $\boldsymbol{\mathcal{D}}$ from $\boldsymbol{\mathcal{D}}^*$, then $\mathcal{B}$ can be used to distinguish $\boldsymbol{Q}$ from $\boldsymbol{Q}^*$, the thing that contradicts DDH assumption for the subgroup $\mathbb{G}$ of matrices generated by $T^{\alpha \zeta_1}$, therefore we conclude that there is no probabilistic polynomial time algorithm $\mathcal{B}$ that distinguishes $K$ from $K^*$ under this assumption.

The argument proceeds as follows. Assume that there is a probabilistic polynomial time algorithm $\mathcal{B}$ that can distinguish between $\boldsymbol{\mathcal{D}}$ and $\boldsymbol{\mathcal{D}}^*$, and assume that $\boldsymbol{Q}$ and $\boldsymbol{Q}^*$ are given. We define the algorithm $\mathcal{H}$ that maps the distribution $\boldsymbol{Q}$ into the distribution $\boldsymbol{\mathcal{D}}$ that simulates the key exchange protocol, and maps $\boldsymbol{Q}^*$ into the distribution $\boldsymbol{\mathcal{D}}^*$ in the same way. Then algorithm $\mathcal{A}$ that distinguishes $\boldsymbol{Q}$ from $\boldsymbol{Q}^*$ is defined as follows.

$\mathcal{A}(\boldsymbol{Q})$:
  (1) Obtain $\boldsymbol{D} = \mathcal{H}(\boldsymbol{Q})$.
  (2) Return $\mathcal{B}(\boldsymbol{D})$.

$\mathcal{A}(\boldsymbol{Q}^*)$:
  (1) Obtain $\boldsymbol{D}^* = \mathcal{H}(\boldsymbol{Q}^*)$.
  (2) Return $\mathcal{B}(\boldsymbol{D}^*)$.

Thus, $Pr[\mathcal{A}(\boldsymbol{Q}) = 1] = Pr[\mathcal{B}(\boldsymbol{D}) = 1]$ and $Pr[\mathcal{A}(\boldsymbol{Q}^*) = 1] = Pr[\mathcal{B}(\boldsymbol{D}^*) = 1]$.

**Algorithm $\mathcal{H}(\mathcal{Q})$:**
Input: $\boldsymbol{Q}(or\ \boldsymbol{Q}^*) = (T^{\alpha\zeta_1}, Y', T^{\gamma\zeta_1}, K'(or\ T^c))$, $M, R, S$.
Output: $\boldsymbol{D}(or\ \boldsymbol{D}^*) = (\boldsymbol{\mu}, \boldsymbol{\sigma}, \boldsymbol{A}, \boldsymbol{B}, Y, K(or\ K^*))$.

1. Select $\boldsymbol{\mu} = (\mu_1,\ldots,\mu_t)$ and $\boldsymbol{\sigma} = (\sigma_1,\ldots,\sigma_t)$ such that
   $\mu_i\ mod\ 2 \neq 0$, $\sigma_i\ mod\ 3 \neq 0$, $(\mu_1 + \cdots + \mu_t) \equiv 0\ mod\ 2p$, and $(\sigma_1 + \cdots + \sigma_t) \equiv 0\ mod\ 3p$.
2. Compute $U = M^3, V = M^2$.
3. For $i = 2,\ldots,t$:
      3.1 Compute $A'_i = (T^{\alpha\zeta_1})^{\zeta'_i} U^{\mu_i}$, $B'_i = (T^{\gamma\zeta_1})^{\zeta'_i} V^{\sigma_i}$
      3.2 Test if $gcd(2\alpha\zeta_1\zeta'_i + \mu_i, 2p) = 1$, $gcd(3\gamma\zeta_1\zeta'_i + \sigma_i, 3p) = 1$, if not, go to step 3.1.
      [This test can be performed testing if none of $A'^2_i, A'^p_i, B'^3_i,$ and $B'^p_i$ is $I_n$].
      3.3 Set $A_i = RA'_i R^{-1}$, $B_i = SB'_i S^{-1}$.
4. Set $\boldsymbol{A} = (A_1, \ldots A_t)$, $\boldsymbol{B} = (B_1, \ldots B_t)$.
5. Set $Y = RY'R^{-1}$, $K = SK'S^{-1}$(or $K^* = ST^c S^{-1}$).
6. Return $\boldsymbol{D}(or\ \boldsymbol{D}^*) = (\boldsymbol{\mu}, \boldsymbol{\sigma}, \boldsymbol{A}, \boldsymbol{B}, Y, K(or\ K^*))$.

Thus, $\mathcal{H}(\boldsymbol{Q})$ returns $\boldsymbol{D}$ and $\mathcal{H}(\boldsymbol{Q}^*)$ returns $\boldsymbol{D}^*$.

One can easily see that the tuple $\boldsymbol{D}$ generated by algorithm $\mathcal{H}$, is statistically the same distribution as $\boldsymbol{D}$ that generated by the key exchange protocol, simply by comparing $\mathcal{H}$ against the key agreement protocol.

Now, if there an algorithm $\mathcal{B}$ such that
$$|Pr[\mathcal{B}(\boldsymbol{D}) = 1] - Pr[\mathcal{B}(\boldsymbol{D}^*) = 1]| > \frac{1}{n^\epsilon}$$
for some $\epsilon > 0$ and arbitrarily large integer $n$.

And since $\mathcal{A}(\boldsymbol{Q})$ has the same return value as $\mathcal{B}(\boldsymbol{D})$, and $\mathcal{A}(\boldsymbol{Q}^*)$ has the same return value as $\mathcal{B}(\boldsymbol{D}^*)$ then
$$|Pr[\mathcal{A}(\boldsymbol{Q}) = 1] - Pr[\mathcal{A}(\boldsymbol{Q}^*) = 1]| > \frac{1}{n^\epsilon}$$
for some $\epsilon > 0$ and arbitrarily large integer $n$.
Which contradicts the DDH assumption for $\mathbb{G}$.
Therefore, there is no probabilistic polynomial time algorithm $\mathcal{B}$ to distinguish between $\boldsymbol{D}$ and $\boldsymbol{D}^*$. Hence, no probabilistic polynomial time algorithm to distinguish $K$ from $K^*$.

## 5 Security Against Quantum Adversary

Consider the description of the protocol $\mathcal{D} = (\boldsymbol{\mu}, \boldsymbol{\sigma}, \boldsymbol{A}, \boldsymbol{B}, Y, K)$, where
$$\boldsymbol{A} = (A_1, \dots, A_t), \quad \boldsymbol{B} = (B_1, \dots, B_t)$$
Note that, since $T = M^6$, $U = M^3$, and $V = M^2$, therefore,
$$A'_i = T^{\alpha \zeta_i} U^{\mu_i} = U^{2\alpha \zeta_i + \mu_i} = M^{3(2\alpha \zeta_i + \mu_i)}$$

And
$$B'_i = T^{\gamma \zeta_i} V^{\sigma_i} = V^{3\gamma \zeta_i + \sigma_i} = M^{2(3\gamma \zeta_i + \sigma_i)}$$

Note also that,
$$\gcd(2\alpha \zeta_i + \mu_i, 2p) = 1, \text{ and } \gcd(3\gamma \zeta_i + \sigma_i, 3p) = 1$$

Thus,
$$A_i = R M^{3(2\alpha \zeta_i + \mu_i)} R^{-1} \tag{5.1}$$
$$B_i = S M^{2(3\gamma \zeta_i + \sigma_i)} S^{-1} \tag{5.2}$$

Suppose that there is a quantum adversary who can solve any type of DLP in polynomial time. To obtain the shared secret key
$$K = B_1^{\theta_1} \dots B_t^{\theta_t} = Y^{\alpha \gamma^{-1}}$$
the adversary must be able to find $\gamma \alpha^{-1}$, or $\theta_1, \dots, \theta_t$. Where these secret numbers exists as exponents in the equations (5.1), (5.2) and
$$Y = A_1^{\theta_1} \dots A_t^{\theta_t} = R M^{6\alpha \sum \zeta_i \theta_i} R^{-1} \tag{5.3}$$

Since the actual generator (i.e. the matrix $M$ generated by algorithm 2.3) used in the protocol is private, the adversary should pick his own generator. However, none of the matrices $A_i$, $B_i$, and $Y$ can be chosen by the adversary as an alternative generator to $M$. Each of $A_i$, $B_i$, and $Y$, has different multiplicative order ($2p$, $3p$ and $p$ respectively), and generates different subgroup, while the multiplicative order of the actual generator $M$ is $\phi = 6p$. Also, for any two integers $x$ and $y$, since $SR \neq RS$, none of the products $A_i^x B_j^y$, or $B_j^x Y^y$ can be used an alternative generator. The product $A_i^x Y^y$ has multiplicative order $2p$, and cannot be chosen as generator.

Moreover, theorem 2 states that $A'^{x}_j \neq B'_i$ for any integer $x$, likewise $B'^{y}_j \neq A'_i$ for any integer $y$, therefore $A_i \neq X B_j^x X^{-1}$ and $B_j \neq Z A_i^y Z^{-1}$ for any matrices $X$ or $Z$. This implies that if $\lambda_{A_i}$ is an eigenvalue of $A_i$ and $\lambda_{B_j}$ eigenvalue of $B_j$, the adversary cannot assume that $\lambda_{A_i} = \lambda_{B_j}^x$, or $\lambda_{B_j} = \lambda_{A_i}^y$.

**Theorem 2**

Given $A'_i = U^{2\alpha\zeta_i + \beta\mu_i}$, $B'_i = V^{3\gamma\zeta_i + \delta\sigma_i}$ (as generated by algorithm 2.2). $A'^x_j \neq B'_i$ for any integer $x$. Likewise $B'^y_j \neq A'_i$ for any integer $y$. $i \in [0, t]$.

**Proof**

For the contradiction, assume that there is $x$ such $A'^x_j = B'_i$ for some $i$ and $j$.
Recall from algorithm 2.2 that order of $A'_j$ is $2p$ and order $B'_i$ is $3p$.
For any $x$ we have only two possibilities: (a) $\gcd(x, 2p) = 1$ and (b) $\gcd(x, 2p) \neq 1$. As we see below, in both possibilities the assumption $A'^x_j = B'_i$ will lead to contradiction.

(a) $\gcd(x, 2p) = 1$. If so, then $\left(A'^x_j\right)^{2p} = I_n = (B'_i)^{2p} \neq I_n$.

(b) $\gcd(x, 2p) \neq 1$. If $\gcd(x, 2p) = 2$, then $x = 2^s r$ and $A'^x_j = A'^{2^s r}_j$ for some integers $r$ and $s$ such that $\gcd(r, 2p) = 1$, thus $\left(A'^x_j\right)^p = \left(A'^{2^s r}_j\right)^p = \left(A'^r_j\right)^{2^s p} = I_n = (B'_i)^p \neq I_n$. Finally, if $\gcd(x, 2p) = p$, then $x = p$ and $\left(A'^x_j\right)^2 = A'^{2p}_j = I_n = B^2_i \neq I_n$.

Therefore there is no $x$ such that $A'^x_j = B'_i$.
$B^y_j \neq A_i$ for any integer $y$ can be proved in the same way.

∎

**What Quantum Adversary Can Do ?**

Since $A'_i = U^{2\alpha\zeta_i + \mu_i}$, $B'_i = V^{3\gamma\zeta_i + \sigma_i}$, there are DLPs

$$A'_j = A'^{(2\alpha\zeta_i + \mu_i)^{-1}(2\alpha\zeta_j + \mu_j)}_i, \quad B'_j = B'^{(3\gamma\zeta_i + \sigma_i)^{-1}(3\gamma\zeta_j + \sigma_j)}_i$$

Form the similarity of $A'_i, B'_i$ and $A_i, B_i$ respectively, we have

$$A_j = A^{(2\alpha\zeta_i + \mu_i)^{-1}(2\alpha\zeta_j + \mu_j)}_i, \quad B_j = B^{(3\gamma\zeta_i + \sigma_i)^{-1}(3\gamma\zeta_j + \sigma_j)}_i$$

Now, assume that the quantum adversary can solve the DLPs $A_i = A^x_j$ and $B_i = B^y_j$ for $x$ and $y$ by solving DLPs between the relevant eigenvalues of these matrices [using the fact that if $\lambda$ is an eigenvalue of matrix $A$, then $\lambda^x$ is an eigenvalue of $A^x$ or using any other technique].
If the adversary picks up $A_j$, let say $A_1$, as a generator for cyclic subgroup that contains $A_2, \ldots, A_t$ and solves the DLPs $A_j = A^{c_j}_1$, $j = 2, \ldots, t$. Then he would come up with the following system of equations:

$$c_j = \alpha\eta_j + \beta\mu_j \quad \text{(I)}$$

where $\eta_j = 2\beta\zeta_j$, $\beta = (2\alpha\zeta_1 + \mu_1)^{-1}$.

Likewise, adversary may picks up $B_1$ as a generator for cyclic subgroup that contains $B_2,\ldots,B_t$ and solves the DLPs $B_j = B_1^{d_j}$, $j = 2,\ldots,t$, to obtain another system of equations:

$$d_j = \gamma\xi_j + \delta\sigma_j \qquad \text{(II)}$$

where $\xi_j = 3\delta\zeta_j$, $\delta = (3\gamma\zeta_1 + \sigma_1)^{-1}$.

The adversary has also these two equation from the description of the protocol.

$$\mu_1\theta_1 + \cdots + \mu_t\theta_t = 2p$$
$$\sigma_1\theta_1 + \cdots + \sigma_t\theta_t = 3p$$

In summary, the best of what a quantum adversary or any adversary from solving DLPs, is the systems of equations (I) and (II) and

$$\left.\begin{array}{l}\mu_1\theta_1 + \cdots + \mu_t\theta_t = 2p \\ \sigma_1\theta_1 + \cdots + \sigma_t\theta_t = 3p\end{array}\right\} \qquad \text{(III)}$$

where $c_j, d_j, p, \mu_j, \sigma_j$ are given positive integers and $\alpha, \beta, \gamma, \delta, \eta_i, \xi_i, \theta_i$ are unknowns. $j \in [1, t]$. To break the protocol the adversary must obtain either $\gamma\alpha^{-1}$ or $\theta_1,\ldots,\theta_t$.

Now, each of (I) and (II) is a system of $(t-1)$ equations in $t+1$ unknowns. And the system (III) is a system of two equations in $t-1$ unknowns. Therefore, when $t > 3$, these systems are unsolvable for $\gamma\alpha^{-1}$ or $\theta_1,\ldots,\theta_t$.
From here we conclude that even the adversary could solve the DLP, the protocol remains secure.

## 6    Conclusion

The proposed scheme aims at overcoming the reliance on the DLP, by hiding the subgroup generator being used as an actual base of exponentiations in the key exchange algorithm. By doing so, we deprive the quantum adversary from advantage of breaking the DLP. Thus breaking this problem would not affect security of the scheme in terms of obtaining the shared secret key computationally. On the other hand, to ensure a standard level of security, we have shown that the scheme is secure against distinguishability of the shared secret key under intractability assumption of Decision Diffie-Hellman Problem.

3. D. Boneh, T. Rsa, R. Rivest, A. Shamir, and L. Adleman, "Twenty Years of Attacks on the RSA Cryptosystem 1 Introduction," Integers, vol. 46, pp. 1–16, 1977.
4. D. Boneh, "The decision diflie-hellman problem," Lect. Notes Comput. Sci. (including Subser. Lect. Notes Artif. Intell. Lect. Notes Bioinformatics), vol. 1423, pp. 48–63, 1998.